\documentclass[showpacs,twocolumn,aps,prl]{revtex4}
\usepackage{amsmath}
\usepackage{amsfonts}
\usepackage{amssymb}
\usepackage{amsthm}
\usepackage{graphicx}

\begin{document}

\title{Current-induced domain wall motion with adiabatic spin torque only in
cylindrical nanowires}
\author{Z. Z. Sun}
\email[Electronic mail: ]{zhouzhou.sun@physik.uni-regensburg.de}
\affiliation{Institute for Theoretical Physics, University of Regensburg,
D-93040 Regensburg, Germany}
\author{J. Schliemann}
\affiliation{Institute for Theoretical Physics, University of Regensburg,
D-93040 Regensburg, Germany}
\author{P. Yan}
\email[Electronic mail: ]{yanpeng@ust.hk}
\affiliation{Physics Department, The Hong Kong University of
Science and Technology, Clear Water Bay, Hong Kong SAR, China}
\author{X. R. Wang}
\affiliation{Physics Department, The Hong Kong University of
Science and Technology, Clear Water Bay, Hong Kong SAR, China}
\date{\today}

\begin{abstract}
We investigate current-driven domain wall (DW) propagation in magnetic
nanowires in the framework of the modified Landau-Lifshitz-Gilbert equation
with both adiabatic and nonadiabatic spin torque (NAST) terms.
Contrary to the common opinion that NAST is indispensable for DW
motion\cite{Klauireview,Thiaville}, we point out
that adiabatic spin torque (AST) only is enough for current-driven
DW motion in a cylindrical (uniaxial) nanowire.
Apart from a discussion of the rigid DW motion from the energy and angular momentum viewpoint,
we also propose an experimental scheme to measure the spin current
polarization by combining both field and current driven DW motion in a flat
(biaxial) wire.
\end{abstract}
\maketitle

The domain wall (DW) motion in magnetic nanowires has
recently attracted much attention in the field of
nanomagnetism\cite{Klauireview}
due to the enormous potential industrial applications\cite{Parkin}, such as
memory bits and logic devices.
Besides the field-induced DW
motion\cite{Ono,Walker,Wang,Sun1}, the current-driven
magnetization reversal
in both magnetic multilayers and nanowires through the spin torque (ST)
transfer mechanism\cite{Slonczewski} has been massively studied,
chiefly under the aspect of
low power consumption and locality of electric currents.
Moreover, a large number of theoretical\cite{Thiaville}
and experimental studies\cite{Klaui,Hayashi} were devoted to the
current-driven DW motion.
Here one distinguishes between the adiabatic spin torque (AST)
which origins from the
polarization of the itinerant electrons adiabatically following the
magnetization direction, and the nonadiabatic spin torque
(NAST, often referred as a $\beta$-term) due to
the mismatch of the current polarization and magnetic moments.
The common current opinion in the literature is that
the latter mechanism is indispensable for current driven DW
motion\cite{Klauireview,Thiaville}.

Experimental studies so far have actually been done on flat
nanowires exhibiting a biaxial magnetic anisotropy rather than
cylindrical wires realizing the uniaxial case.
This is mainly due to difficulties to produce cylindrical metallic wires
by conventional lithography although there are techniques available to
grow them via electrodeposition\cite{Fert}.
In this letter, starting from an analytical analysis of the
modified Landau-Lifshitz-Gilbert (LLG) equation, we show that
AST only is enough to drive a sustained DW motion in a cylindrical wire.
Our result relies on the fact that the magnetic moments in a cylindrical wire
precess around the easy axis, while they rotate in a plane in the
biaxial case\cite{Walker}. Furthermore, in the uniaxial case
no Walker breakdown occurs for sufficiently large currents (or fields), as also
shown recently by  micromagnetic simulations\cite{Yan}.
Finally, we also propose a scheme to measure the spin current
polarization $P$ by combining both field and current driven DW motion.

A magnetic nanowire can be described as an effectively one-dimensional
(1D) continuum of magnetic
moments along the wire axis direction. Magnetic domains are formed due
to the competition between the anisotropic magnetic energy and the exchange
interaction among adjacent magnetic moments.
Without loss of generality, we shall consider a head-to-head DW structure,
assuming the easy axis being along the wire ($z$-) axis.
A biaxial anisotropy energy density can be formulated as
$\varepsilon^K =-KM_z^2+K'M_x^2$
where $K$ and $K'$ describe the easy and hard axis anisotropy
pointing along the $z$-axis and the $x$-axis, respectively.
The case $K'=0$ describes a uniaxial anisotropy along the wire axis.
The above energy density is given in units of
$\mu_0M_s^2$, where $\mu_0$ is
the vacuum permeability, and $M_s=|\vec{M}|$ such that the local
magnetization $\vec{M}$ enters here as a unit vector,
$\vec{m}=\vec{M}/M_s$.

If a current is driven through the nanowire, the
spatio-temporal magnetization dynamics is governed by the so-called modified
LLG equation with additional AST and NAST terms\cite{Thiaville},
\begin{equation}
\dot{\vec{m}}= -|\gamma|\vec{m}\times\vec{H}_{t}
+\alpha \vec{m}\times
\frac{\partial \vec{m}}{\partial t}-(\vec{u}\cdot\nabla)\vec{m}
+\beta \vec{m}\times[(\vec{u}\cdot\nabla)\vec{m}],\label{LLG}
\end{equation}
where $|\gamma|$, $\alpha$, and $\beta$ are the gyromagnetic ratio,
the Gilbert damping coefficient, and a dimensionless coefficient
describing the NAST strength, respectively. $\beta$ is usually of
the same order as the damping in ferromagnetic metals.
The velocity $\vec{u}$ points along the flow direction
of the itinerant electrons which is usually the wire axis although
perpendicular current injection has also been proposed\cite{Khvalkovskiy}.
Thus, $\nabla=\partial /\partial z$ in 1D, and $u=g\mu_B JP/(2eM_s)$
where (among standard notation)
$J$ and $P$ are the density and the spin polarization
of the current, respectively.

The (normalized) effective field $\vec{h}_t\equiv \vec{H}_{t}/M_s$ is given by
the variational derivative of the total energy density (per unit section-area)
$E=\int_{-\infty}^{\infty}dz\varepsilon(z)$ with respect to magnetization,
$\vec{h}_{t}=-\delta E/ \delta \vec{m}(z)$. The local energy density
is given by\cite{Sun1}
\begin{equation}
\varepsilon(z)=
-K m_z^2 +K' m_x^2 +A(\theta'^2
+\sin^2\theta \phi'^2)
-  \vec{m} \cdot \vec{h},\label{energy}
\end{equation}
where $A$ describes the exchange interaction, and $\vec{h}$ is the normalized
external field. Moreover, we have adopted the usual spherical coordinates,
$\vec{m}(z,t)=(\sin\theta\cos\phi, \sin\theta\sin\phi, \cos\theta)$
where the  polar angle $\theta(z,t)$ and the azimuthal angle $\phi(z,t)$
depend on position and time, and the prime denotes spatial differentiation.

Following Ref.~\cite{Walker,Sun1}, we will focus on DW structures fulfilling
$\phi'=\phi''=0$, i.e. all the magnetic moments synchronously rotate around
the easy axis in space. Then the dynamical LLG equations take the form
\begin{align}
&\Gamma\dot{\theta} = \alpha (2A\theta''-K\sin 2\theta
-K'\sin2\theta\cos^2\phi ) \nonumber\\
&+K'\sin\theta\sin2\phi+(\alpha h_{\theta}+h_{\phi})-(1+\alpha\beta)u\theta',
\nonumber\\
&\Gamma\sin\theta\dot{\phi} = -(2A\theta''-K\sin 2\theta
-K'\sin2\theta\cos^2\phi )\nonumber\\
&+\alpha K'\sin\theta\sin2\phi+(\alpha h_{\phi}-h_{\theta})-(\alpha-\beta)u\theta', \label{llg1}
\end{align}
where we have defined $\Gamma \equiv 1+\alpha^2$ and introduced
a dimensionless time via $t\mapsto t|\gamma| M_s$.
$h_i (i=r,\theta,\phi)$ are the components of the external field in
spherical coordinates.


The linear DW motion under field or ST in a biaxial wire has already been discussed in the literature\cite{Walker, Thiaville}.
Let us first review the case of a flat wire with biaxial anisotropy.
Following the pioneering work by Schryer and Walker\cite{Walker} we
concentrate on solutions fulfilling $\phi(z,t)\equiv\phi_0={\rm constant}$.
This assumption implies all the magnetic moments move in a plane and is valid
at sufficiently low fields\cite{Walker}.
Substituting the travelling-wave ansatz
$\tan \frac{\theta}{2} =\exp\left(\frac{z-vt}{\Delta}\right)$
into Eqs.~\eqref{llg1}, we obtain
\begin{align}
& \Gamma v= \Delta(\alpha h-K'\sin 2\phi_0)+(1+\alpha\beta)u\,,\nonumber\\
& \Delta( \alpha K' \sin 2\phi_0 + h)-(\alpha-\beta)u=0\,,\nonumber
\end{align}
where $\Delta\equiv\sqrt{A/(K +K' \cos^2\phi_0)}$ is the DW width, and the external field with a magnitude $h$ along the $z$ axis is assumed ($h_{\theta}=-h \sin\theta,h_{\phi}=0$). Thus, the constant plane angle $\phi_0$ and the DW
velocity satisfy
\begin{equation}
\alpha K' \sin2\phi_0=(\alpha-\beta)u/\Delta-h, \quad v=\Delta h/\alpha +\beta u/\alpha. \label{biv}
\end{equation}
This solution just recovers the Schryer-Walker result\cite{Walker}
in the presence of
ST and implies $\beta\neq 0$ i.e. NAST is indispensable for nonzero
DW velocity\cite{Thiaville}.
In particular, for $|(\alpha-\beta)u/\Delta- h|>\alpha K'$
the sine in Eq.~\eqref{biv} becomes larger than unity, and the
solution
breaks down. Therefore, the limit of uniaxial anisotropy
$K'\to 0$ cannot be reached within the above travelling-wave ansatz.

However, the results~\eqref{biv},
devise the following scheme to experimentally determine the spin
polarization of the current combining measurements of field-driven and
current-driven DW dynamics:
First perform velocity measurement using field- and current-driven
DWs separately and obtain the quantities
$\Delta_{min}/\alpha\equiv C_1$ and $\beta P/\alpha\equiv C_2$. Here
$\Delta_{min}=\sqrt{A/(K+K')}$ is the minimum DW width and $C_1$ can
be obtained by extrapolating the data to $h\to 0$. Then apply
a fixed field such that the Walker limit is reached and the
DW width reaches its maximum, $\Delta_{max}=\sqrt{A/(K+K'/2)}$.
By injecting a spin-polarized current and subsequently
lowering the current density and monitoring the decrease of
DW width one can reach the situation
$(\alpha-\beta)u=\Delta h$ implying $\sin2\phi_0=0$ and
again $\Delta=\Delta_{min}$.
(If $\beta>\alpha$ one may reverse the injected current direction.)
Now using $(\alpha-\beta)P/\Delta_{min}\equiv C_3$
one can infer the current polarization
$P=C_1C_3+C_2$ and $\beta/\alpha=C_2/(C_1C_3+C_2)$.
A large anisotropy i.e. $K'\gg K$ and the resolution for observing the DW
width variation are the key points for this scheme.

Now we look at the DW motion in a uniaxial wire ($K'=0$) which is essentially
different from the biaxial case. Using the travelling-wave ansatz\cite{Sun1} $\tan \frac{\theta}{2} =\exp\left(\frac{z-vt}{\Delta}\right)$ where now
$\Delta =\sqrt{A/K}$, and also for a static field $h$ applied along the $z$ axis, one can straightforwardly obtain
the following expressions
for the velocity and the
precession frequency from Eq.\eqref{llg1},
\begin{equation}
\Gamma v=\alpha \Delta h+(1+\alpha\beta)u, \qquad \Gamma\dot{\phi} = h-(\alpha-\beta)u/\Delta. \label{univ}
\end{equation}

From Eq.~\eqref{univ}, we immediately conclude an important result
that sustained DW motion is possible in a cylindrical nanowire
even in the absence of NAST, a finding contradicting leading opinions in the
recent literature. Moreover, Eq.~\eqref{univ} also shows that, at small
damping, the velocity of a current-driven DW is essentially proportional to
$u$ while in the field-driven case we have $v\approx\alpha\Delta h$
and the DW motion is suppressed. Thus, in a uniaxial nanowire, a current is
more efficient in driving a DW than an external field.
Furthermore, the precessional frequency also occurs in the recently
studied DW driven electromotive force $V_{emf}=\pm \hbar\dot{\phi}/e$
\cite{Barnes1}, which should also motivate future experiments.
In particular, in a cylindrical nanowire the current polarization
can be easily determined experimentally by measuring the velocity of a
current-driven DW.

Let us now discuss the energy variation in the field and/or current driven case
in both types of nanowires. In general, the change rate of total magnetic
energy (from anisotropy, exchange interaction and external magnetic field)
can be expressed as
\begin{eqnarray}
\dot{E} & = & -\int_{-\infty}^{\infty}dz \dot{\vec{m}}
\cdot [\alpha \dot{\vec{m}}+u(\vec{m}\times\partial_z\vec{m})
+\beta u \partial_z\vec{m}]\nonumber\\
 & \equiv & P_{\alpha}+P_{\rm AST}+P_{\rm NAST}\,. \label{power}
\end{eqnarray}
Here $P_{\alpha}=-2\alpha\Delta(\dot{\phi}^2+v^2/\Delta^2)$
is the total dissipation power due to damping, and
$P_{\rm AST}=-2u\dot{\phi}$ and $P_{\rm NAST}=2\beta u v/\Delta$ are the energy
pumping rates induced by AST and NAST, respectively.
Note that the DW width $\Delta$ follows different expressions in the
biaxial and uniaxial case, and $v$ and $\dot{\phi}$ are given by
Eqs.~\eqref{biv} and \eqref{univ}.
Remarkably, the AST does not contribute to any change in energy
for a biaxial wire ($\dot{\phi}=0$) below the Walker limit. Moreover, it is straightforward to verify
that in both cases it holds $\dot{E}=-2hv$ which is just the released Zeeman
energy per time of a rigid DW traveling with velocity $v$ in the
external field $h$\cite{Wang}. Thus, the relation $\dot{E}=-2hv$ can also be
used as a definition of the DW velocity from an energetic point of view.
In particular, a moving DW driven
purely by current (zero magnetic field) conserves its energy.


Let us now consider the change of angular momentum
$\vec{L}=\int dz \vec{m}(z,t)$ (per section-area and in units of
$-\mu_0M_s/|\gamma|$) due to ST. In one hand, the angular momentum
change due to rigid DW motion is $\dot{\vec{L}}=\pi\dot{\phi}\Delta \hat{e}_{\phi} +2v\hat{z}$. On the other hand, LLG equation \eqref{LLG} gives
\begin{eqnarray}
\dot{\vec{L}} & = & \int_{-\infty}^{\infty}dz [\vec{h}_t \times \dot{\vec{m}}
+\alpha \vec{m} \times \dot{\vec{m}} -u\partial_z\vec{m}
-\beta u \vec{m}\times\partial_z\vec{m}]\nonumber\\
 & \equiv & \vec{\Gamma}_{pre}+\vec{\Gamma}_{\alpha}
+\vec{\Gamma}_{\rm AST}+\vec{\Gamma}_{\rm NAST}\,.
\end{eqnarray}
For our uniaxial wire,
$\vec{\Gamma}_{pre}=\pi h\Delta\hat{e}_{\phi}$,
$\vec{\Gamma}_{\alpha}=-\pi\alpha v\hat{e}_{\phi}+2\alpha \dot{\phi}\Delta \hat{z}$,
$\vec{\Gamma}_{\rm AST}=2u\hat{z}$
and $\vec{\Gamma}_{\rm NAST}=\pi\beta u\hat{e}_{\phi}$.
Hence, the AST pumps a longitudinal spin to the wire to push the DW
propagating while the NAST and external field provide the transverse
torques forcing DW precession around the wire axis. Furthermore, $\vec{\Gamma}_{\alpha}$ due to the DW precession ($\dot{\phi}$)
provides an extra longitudinal torque and that due to the propagation ($v$)
results in an effective transverse force. These two damping effects are
reminiscent of Barnett effect\cite{Barnett} and Einstein-de Haas
effect\cite{Einstein}, respectively. In the present case, both
effects originate from non-zero damping $\alpha\neq 0$.
Moreover, two of us have recently studied DW motion driven by a circularly
polarized microwave, which also embodies the Barnett effect\cite{Yanpeng}.
We also note that Eqs.~\eqref{univ} can equivalently expressed as
\begin{equation}
v=u+\alpha \dot{\phi} \Delta, \qquad \dot{\phi}=h+(\beta u-\alpha v)/\Delta,
\end{equation}
which shows explicitly that the DW precession contributes to its velocity
as $\alpha \dot{\phi} \Delta$,  and the DW translation contributes to
the precession as $-\alpha v/\Delta$.
The above considerations refer to a uniaxial wire. In the biaxial case,
however, $\vec{\Gamma}_{pre}$ acquires an additional
term from the transverse anisotropy $K'$ and the DW precession is
absent. As a result, the effect of AST is canceled and only NAST affects
the DW velocity.

Before ending the letter, we would like to give a practical example of the
DW motion in a permalloy uniaxial or biaxial nanowire. We employ
the standard data $M_s=800$kA/m, $\Delta=20$nm, $\alpha=0.02,
\beta=0.04, P=0.4$\cite{Thiaville, Ono}. For a uniaxial wire,
$v$(m/s)$\approx 0.7H$(100Oe)$ + 2.9J$(10$^{7}$A/cm$^2$), $\dot{\phi}$(GHz)
$\approx 1.76H$(100Oe)$+0.003J$(10$^7$A/cm$^2$); for a flat wire,
$v$(m/s)$\approx 1760H$(100Oe)$+5.8J$(10$^7$A/cm$^2$) below the Walker limit.
Thus the current-induced DW velocity has the same order in both
 types of wires. The field-driven azimuthal precessional frequency in a
cylindrical wire is in GHz regime while current-driven precessional
frequency is in MHz regime.


In summary, we have demonstrated, in the framework of the modified
Landau-Lifshitz-Gilbert equation,
that adiabatic spin torque only is enough for current-driven
DW motion in a cylindrical (uniaxial) nanowire. This finding
contradicts leading opinions expressed in the recent literature.
Moreover, we have also discussed the motion of a rigid DW being
subject to (adiabatic or nonadiabatic) spin torque
in flat or cylindrical wires from an energetic and angular momentum point of view.
Finally, we have also proposed an experimental scheme to measure the spin
current
polarization by combining both field and current driven DW motion in a flat
(biaxial) wire.

Z.Z.S. thanks the Alexander von
Humboldt Foundation (Germany) for a grant.
This work has been supported by Deutsche Forschugsgemeinschaft via SFB 689.
PY and XRW are supported by Hong Kong RGC grants (604109 and HKU10/CRF/08-
HKUST17/CRF/08).

\end{document}